\documentclass[final]{IEEEtran}

\ifCLASSINFOpdf
\else
   \usepackage[dvips]{graphicx}
\fi
\usepackage{url}
\usepackage{graphicx}
\usepackage{placeins}
\usepackage{cite}
\usepackage{float}
\usepackage{bbm}
\usepackage{subfigure}
\usepackage{verbatim}
\usepackage{amssymb}
\usepackage{amsmath}
\usepackage{amsthm}
\usepackage[mathscr]{euscript}
\usepackage[compact]{titlesec}
\usepackage{bbm}
\usepackage{mwe}
\usepackage[ruled,vlined]{algorithm2e}

\usepackage{graphicx}

\DeclareMathOperator{\E}{\mathbb{E}}

\DeclareMathOperator*{\argmin}{argmin}

\newtheorem{definition}{Definition}
\newtheorem{remark}{Remark}

\usepackage{setspace}

\titlespacing{\section}{4pt}{0.25ex}{0.5ex}
\titlespacing{\subsection}{0pt}{0.25ex}{0.5ex}
\titlespacing{\subsubsection}{0pt}{0.25ex}{0.5ex}

\begin{document}

\title{Constrained Online Learning to Mitigate Distortion Effects in Pulse-Agile Cognitive Radar}
\author{Charles E. Thornton, R. Michael Buehrer, and Anthony F. Martone
\thanks{C.E. Thornton and R.M. Buehrer are with Wireless @ VT, Bradley Department of ECE, Virginia Tech, Blacksburg, VA, 24061, USA. \textit{(e-mails:$\{$thorntonc, buehrer$\}$@vt.edu)}.}
\thanks{A.F. Martone is with the US Army Research Laboratory, Adelphi, MD, 20783, USA. \textit{(e-mail: anthony.f.martone.civ@mail.mil).}}
\thanks{The support of the US Army Research Office is gratefully acknowledged.}}

\markboth{}
{Constrained Online Learning to Mitigate Distortion Effects in Pulse-Agile Cognitive Radar}
\maketitle
\begin{abstract}
Pulse-agile radar systems have demonstrated favorable performance in dynamic electromagnetic scenarios. However, the use of non-identical waveforms within a radar's coherent processing interval may lead to harmful distortion effects when pulse-Doppler processing is used. This paper presents an online learning framework to optimize detection performance while mitigating harmful sidelobe levels. The radar waveform selection process is formulated as a linear contextual bandit problem, within which waveform adaptations which exceed a tolerable level of expected distortion are eliminated. The constrained online learning approach is effective and computationally feasible, evidenced by simulations in a radar-communication coexistence scenario and in the presence of intentional adaptive jamming. This approach is applied to both stochastic and adversarial contextual bandit learning models and the detection performance in dynamic scenarios is evaluated.
\end{abstract}

\IEEEpeerreviewmaketitle
\vspace{-.1cm}
\section{Introduction}
\begin{figure}[t]
	\centering
	\includegraphics[scale=0.5]{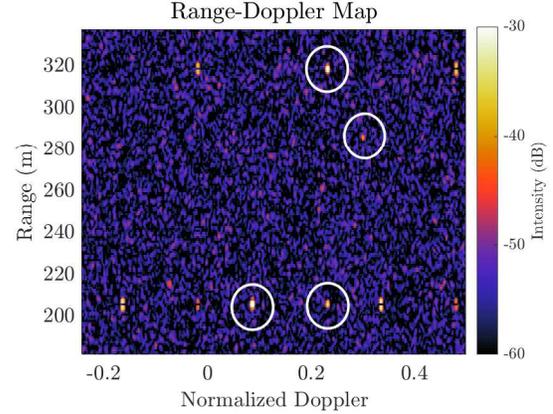}
	\caption{An example of harmful distortion effects due to pulse agility. The four true targets are located in white circles. Other features are due to distortion effects and noise.}
	\label{fig:traj}
\end{figure}

To meet the strict performance and interoperability demands of modern sensing applications, a large body of work on \textit{cognitive radar} (CR) has emerged \cite{aesmag,ontheroad,practical,bell15}. CR aims to improve the radar's overall awareness by using closed-loop feedback between the transmitter and receiver to learn fundamental characteristics of the environment and optimize sensor-processor parameters accordingly. In time-varying applications, such as spectrum sharing, target tracking, or electronic warfare, CR may require pulse-to-pulse agility in its waveform selection process \cite{kirk20,dqnradar,experimentalNotch}. However, an unintended consequence of employing non-identical waveforms within a Coherent Processing Interval (CPI) is the potential target distortion effects and clutter modulation in the received data matrix when coherent range-Doppler processing is applied using a matched or mismatched filter \cite{blunt16}. An example of these unintended effects can be seen in Figure \ref{fig:traj}. As a result, CR schemes must account for the potentially hazardous effects of intra-CPI waveform adaptations to improve overall awareness.

Several contributions have aimed to mitigate the distortion and clutter modulation effects that follow from pulse agility using adaptive processing techniques. In \cite{richardson}, Richardson-Lucy deconvolution, an iterative technique to deblur an image corrupted by a known point-spread function, is applied to range-Doppler maps to mitigate distortion effects. In \cite{nimpc} and subsequently \cite{ravenscroft}, a joint range-Doppler processing technique is developed and applied to pulse-agile cognitive radar transmissions, demonstrating an improvement over both matched and mismatched filtering. Similarly, \cite{scholnik11} develops a framework for clutter cancellation of non-identical pulses and demonstrates that similar filtering techniques can be applied to a broad class of waveforms. Frequency diverse array processing has also been proposed to accommodate pulse-diverse waveforms in MIMO radar \cite{fdamimo}.

While the aforementioned processing techniques are capable of reducing the number of false alarms due to distortion effects, the computational cost is often very high and scales with the size of the received data matrix \cite{richardson},\cite{nimpc}. As CR matures to incorporate more holistic situational awareness, it becomes both practical and necessary to address distortion effects while optimizing parameters within the CR transmitter-receiver feedback loop, which is the subject of this study.

\textit{Contributions:} This work develops an online learning framework to optimize CR waveform parameters in an unknown, time-varying, spectral environment. Specific waveform adaptations with a high likelihood of distortion are eliminated by the learning algorithm. In simulated dynamic scenarios, the present learning approach is shown to provide favorable detection performance compared to unconstrained learning and traditional radar operation. To the best of our knowledge, this is the first work to directly address distortion effects due to non-identical radar pulses from the perspective of the radar transmitter.

\textit{Notation:} The following notation is used. Bold upper (and lower) case letters denote matrices (and vectors) $\mathbf{X}$ (and $\mathbf{x}$). $\mathbf{X}^{T}$ is the transpose operation. Upper case script letters, such as $\mathcal{A}$, denote sets. $\mathbf{I}_{d}$ is the $d \times d$ identity matrix. $\mathbf{0}_{d}$ is a length $d$ vector of zeros. $\langle \cdot, \cdot \rangle$ is the inner product operation. $\lVert \cdot \rVert$ is the $\ell_{2}$-norm. $\mathbb{P}(\cdot)$ is a probability measure. $\mathbb{R}$ and $\mathbb{N}$ denote the sets of all real and natural numbers, respectively. $\mathbbm{1}\{\cdot\}$ is the indicator function, which returns $1$ if the argument is true and $0$ otherwise.

\section{Problem Formulation}
Consider a stationary, monostatic, CR system located at the origin. Time is slotted into a sequence of discrete intervals $\{t\}$, where each index $t \in \mathbb{N}$ corresponds to the CR's $t^{\text{th}}$ pulse repetition interval (PRI). Each PRI, a linear frequency modulated (LFM) chirp waveform $w_{i}$ must be selected from a finite catalog $\mathcal{W} = \{ w_{i} \}_{i = 1}^{W}$. The transmitted waveform is given by \cite{skolnik}
\begin{equation}
w_{i}(t^{\prime}) = A \; \texttt{rect}(t^{\prime} / T) \cos{(2 \pi f_{c,i} t^{\prime} + \pi \alpha_{i} t^{\prime 2})},
\end{equation}
where $t^{\prime}$ corresponds to continuous \textit{`fast time'} between PRI's, $A$ is a constant amplitude, $T$ is the pulse time, $f_{c,i}$ is the center frequency, and $\alpha_{i}$ is the slope of the up-chirp frequency, which dictates the signal bandwidth, given by $\texttt{BW}_{i} = T \alpha_{i}$. The signal is transmitted over a wireless propagation channel that contains one or more targets and may also contain some interference, noise, and clutter. The received signal is then given by
\begin{multline}
y_{i}(t^{\prime}) = \textstyle \sum_{k = 1}^{N_{\texttt{tar}}} g_{k} w_{i}(t^{\prime}-\tau_{k}) \exp(-j v_{k} f_{c} (t^{\prime} - \tau_{k}) / c) \\ + \tilde{i}(t^{\prime}) + \tilde{c}(t^{\prime}) + \tilde{n}(t^{\prime}),
\end{multline}
where $N_{\texttt{tar}}$ is the number of targets, $g_{k}$, $\tau_{k}$, and $v_{k}$ are the gain, round-trip delay, and velocity due to target $k$, $c$ is the speed of light, and $\tilde{i}$, $\tilde{c}$, $\tilde{n}$ are interference, clutter, and noise terms respectively.

To aid in waveform selection, it is assumed the radar capable of passively sensing the spectrum in a shared channel during each PRI \cite{Martone2018}. The spectrum sensing process yields an estimated interference vector $\mathbf{\hat{s}}_{t} = [s_{1},...,s_{S}]$, which contains information about the interference power in a fixed number $S \in \mathbb{N}$ of sub-channels of predetermined size. Each vector element $s_{i} \in \{0,1\}$ is a binary value,\footnote{The total number of unique values the interference vector can take is thus $2^{S}$} where zero denotes that the average interference power in band $i$ is below harmful threshold $H$ and one corresponds to average interference power above $H$ in band $i$. The set of possible interference vectors is denoted by $\mathscr{S}$. In general, it is not guaranteed that the interference will remain stationary in the time between passive spectrum sensing and waveform selection. Thus, the estimated interference vector $\mathbf{\hat{s}}_{t}$ can be thought of as \textit{side information} which is an estimate of some true interference vector $\mathbf{s}_{t} \in \mathscr{S}$. The bandwidth occupied by interference $\mathbf{s}_{t}$ is denoted by $\texttt{BW}_{\mathbf{s}_{t}}$ and the center frequency is denoted by $f_{c,\mathbf{s}_{t}}$.

Each PRI, the CR wishes to solve the optimization problem $w_{t}^{*} = \argmin_{w_{i} \in \mathcal{W}} \E[\mathcal{C}(w_{i},\mathbf{s}_{t})|\mathbf{\hat{s}_{t}}, \mathcal{F}_{t-1}]$, where $\mathcal{C}: \mathcal{W} \times \mathscr{S} \mapsto [0,1]$ is a function which decides the relative \emph{cost} associated with transmitting $w_{i} \in \mathcal{W}$ when the true interference vector is $\mathbf{s}_{t}$. $\mathcal{F}_{t-1}$ is the $\sigma$-algebra generated from the history of transmitted waveforms, interference vector estimates, and observed costs until PRI $t-1$. It is assumed that the radar can store $\mathcal{F}_{t-1}$ in memory each PRI to enable learning over time. A description of the cost function requires the following definitions.
\begin{definition}
	Define the collision bandwidth as 
	\begin{multline}
	\texttt{BW}_{c}(w_{i},\mathbf{s}_{t}) \triangleq \frac{1}{S} \sum_{j = 1}^{S} ( \mathbbm{1}\{j \in [f_{c,i}-\texttt{BW}_{i},f_{c,i}+\texttt{BW}_{i}]\} \times \\ \mathbbm{1}\{j \in [f_{c,\mathbf{s}_{t}}-\texttt{BW}_{\mathbf{s}_{t}},f_{c,\mathbf{s}_{t}}+\texttt{BW}_{\mathbf{s}_{t}}] \} ), 
	\end{multline}
	which corresponds to the fraction of the shared channel bandwidth occupied by both the radar's waveform $w_{i}$ and the interference state vector $\mathbf{s}_{t}$. 	
\end{definition}
\begin{definition}
	Define the missed bandwidth as
	\begin{equation}
	\texttt{BW}_{miss}(w_{i},\mathbf{s}_{t}) \triangleq \frac{\texttt{BW}_{i'} - \texttt{BW}_{i}}{B},
	\end{equation}
	where $B$ is the total bandwidth of the shared channel and $BW_{i'}$ is the bandwidth of waveform $w_{i'} \in \mathcal{W}$ that utilizes the largest bandwidth out of the waveforms that have zero collision bandwidth with $\mathbf{s}_{t}$. If no waveforms in $\mathcal{W}$ have zero collision bandwidth with $\mathbf{s}_{t}$, $BW_{miss} = 0$.
\end{definition}
\begin{definition}
	Let the distortion function be
	\begin{equation}
	 \mathcal{D}(w_{t}|\mathcal{F}_{t-1}) \triangleq \gamma_{1} \lVert f_{c,t} - f_{c,t-1} \rVert^{2} + \; \gamma_{2} \lVert \texttt{BW}_{t} - \texttt{BW}_{t-1} \rVert^{2} 
	 \end{equation}
	 where $f_{c,t}, f_{c,t-1}$ and $\texttt{BW}_{t}, \texttt{BW}_{t-1}$ are the respective center frequencies and bandwidths of $w_{t}$ and $w_{t-1}$. $0 \leq \gamma_{1}, \gamma_{2} \leq 1$ are parameters such that $\mathcal{D}(w_{t}|\mathcal{F}_{t-1}) \in [0,1]$.
\end{definition}
\begin{definition}
Let the CR's cost function be
\begin{equation}
 \mathcal{C}(w_{i},\mathbf{s}_{t}) \triangleq \beta_{1} \texttt{BW}_{c} + \beta_{2} \texttt{BW}_{miss} + \beta_{3} \mathcal{D}(w_{i}|\mathcal{F}_{t-1}) 
 \end{equation}
 where $0 \leq \beta_{1},\beta_{2},\beta_{3} \leq 1$ are parameters such that $\beta_{1}+\beta_{2}+\beta_{3} = 1$. To calculate the cost, it is assumed the radar can recover $\mathbf{s}_{t}$ from received signal $y_{i}$ via \texttt{SINR} estimation.
\end{definition}
\begin{remark}
The CR cost function $\mathcal{C}(w_{i},\mathbf{s}_{t})$ is bounded $ \in [0,1]$. Further, the cost function is locally Lipschitz continuous, meaning that for any two waveforms $w_{j} \in \mathcal{W}$ and $w_{k} \in \mathcal{W}$, $\lvert \mathcal{C}(w_{j}) - \mathcal{C}(w_{k}) \rvert \leq \mathcal{L}(w_{j},w_{k})$, where $\mathcal{L}$ is a metric of distance between waveforms. Thus, the waveform selection set forms a \emph{metric space}, which allows for similarity between waveforms to be exploited by online learning algorithms.
\end{remark}
\begin{proof} \renewcommand{\qedsymbol}{}
See Appendix A.
\end{proof}
By selecting waveforms which yield low average cost, the CR is equivalently attempting to minimize the cumulative \textit{strong regret} experienced in period $\mathcal{T}$, defined by
\begin{equation}
\texttt{Regret}(\mathcal{T}) \triangleq \textstyle \sum_{t = 1}^{\mathcal{T}} \left[ \mathcal{C}({w_{t}^{*}},\mathbf{s}_{t}) - \mathcal{C}({w_{t}},\mathbf{s}_{t}) \right],
\label{eq:regret} 
\end{equation}
where $w_{t}^{*}$ is the waveform which minimizes $\mathcal{C}(\cdot,\mathbf{s}_{t})$ in PRI $t$ and $w_{t}$ is the waveform transmitted by the CR in PRI $t$. Since calculation of (\ref{eq:regret}) requires knowledge of $w_{t}^{*}$ at each step, it must be calculated in hindsight. Thus, the strong regret cannot be minimized directly, and online optimization must be used to select actions in each PRI such that $w_{t}^{*}$ is selected by the radar as often as possible in expectation.

\section{Online Learning Framework}
To select the cost-optimal waveform, the CR must estimate $\E[\mathcal{C}(w_{i},\mathbf{s}_{t})|\mathbf{\hat{s}}_{t}, \mathcal{F}_{t-1}]$ for each pair $(w_{i}, \mathbf{\hat{s}}_{t}) \in \mathcal{W} \times \mathscr{S}$. Thus, a natural trade-off between \textit{exploration} and \textit{exploitation} arises. Each waveform must be transmitted enough times in different interference contexts to reliably predict the expected cost using $\mathcal{F}_{t-1}$, while the total number of sub-optimal waveforms transmitted in period $\mathcal{T}$ should be minimized. To balance exploration and exploitation, the problem is formulated using both stochastic and adversarial linear contextual bandit models. To address concerns regarding distortion effects, both schemes apply a constrained optimization criterion based on the distortion metric.

\subsection{Stochastic Linear Contextual Bandits and Thompson Sampling}
We first study a stochastic linear contextual bandit learning model, under which the cost at each PRI is characterized by the following structure
\begin{equation}
\mathcal{C}(w_{i}, \mathbf{s}_{t}) = \langle \boldsymbol{\theta}, \mathbf{x}_{w_{i},t} \rangle + \eta_{t},
\label{eq:stochmodel}
\end{equation}
where $\boldsymbol{\theta} \in \mathbb{R}^{d}$ is a parameter vector that the radar wishes to learn, $\mathbf{x}_{w_{i},t} \in \mathbb{R}^{d}$ is a \emph{context} vector, associated with each waveform $w_{i}$ at time $t$, which is assembled using information about previous contexts, transmitted waveforms and costs from the $\sigma$-algebra $\mathcal{F}_{t-1}$, which is stored in memory. In this implementation, the context vector contains the following features, 
\begin{equation*}
\begin{aligned}
\xi_{1} = \bar{\mathcal{C}}(w_{i},\mathbf{\hat{s}}_{t}), \; \; & \xi_{2} = \frac{\sum_{\ell}(\mathcal{C}_{\ell}(w_{i},\mathbf{\hat{s}}_{t})-\bar{\mathcal{C}}(w_{i},\mathbf{\hat{s}}_{t}))^{2}}{N_{c}-1},\\ & \text{and} \; \; \xi_{3} = \mathcal{C}_{N_{c}}(w_{i},\mathbf{\hat{s}}_{t}),
\end{aligned}
\end{equation*}
where $N_{c}$ is the number of times the context-action pair $(w_{i},\mathbf{\hat{s}}_{t})$ has been encountered. $\xi_{1}$ is the sample mean of all observed $C(w_{i},\mathbf{\hat{s}}_{t})$ instances where the sum is taken over a sub $\sigma$-algebra of $\mathcal{F}_{t-1}$ that contains only instances of the context-action pair of interest. $\xi_{2}$ is the sample variance of costs over the same sub $\sigma$-algebra containing instances of $\mathcal{C}(w_{i},\mathbf{\hat{s}}_{t})$. $\xi_{3}$ is the most recently observed instance of $\mathcal{C}(w_{i},\mathbf{\hat{s}}_{t})$. 

Returning to the description of (\ref{eq:stochmodel}), $\eta_{t}$ is a noise term, which reflects cases in which the inner product $\langle \boldsymbol{\theta}, \mathbf{x}_{w_{i},t} \rangle$ does not explicitly predict the cost due to fluctuations in the environment or estimation errors. The linear relationship between the context vectors and the costs through an inner product with $\boldsymbol{\theta}$ allows for learning to transfer between contexts, which is a powerful tool when particular contexts may occur infrequently.

It is assumed that $\eta_{t}$ is conditionally 1-subgaussian \cite{lattimore}, which precisely means that for any $\lambda \in \mathbb{R}$
\begin{equation}
	\E[\exp(\lambda \eta_{t}) |\mathcal{F}_{t}] \leq \exp\left(\frac{\lambda^{2}}{2}\right) \; \; \; \textit{almost surely},
\end{equation}
which implies that $\eta_{t}$ has a tail that decays faster than a Gaussian distribution. Practically, this means that there exists a parameter vector $\boldsymbol{\theta}$ such that knowledge of $\boldsymbol{\theta}$ will allow the radar to select the waveform with the lowest expected cost very often.

Previous work in radar and communications has found the stochastic model to be viable for many wireless problems due to the underlying randomness of the channel \cite{dsa1,thornton,Amuru2016}. Many efficient algorithms have been well-studied in the stochastic setting, such as upper confidence bound and $\epsilon$-greedy strategies \cite{lattimore}. However, a Bayesian inspired heuristic called Thompson Sampling (TS) has attracted significant attention in the online learning literature due to near optimal empirical performance on a variety of tasks, which has recently been by supplemented by theoretical results offering favorable performance guarantees \cite{agrawal}, \cite{infoTS}.
\begin{algorithm}[t]
	\setlength{\textfloatsep}{0pt}
	\label{algo:ts}
	\caption{Constrained Linear Contextual Thompson Sampling}
	\SetAlgoLined
	Initialize parameters $\mathbf{B} = \mathbf{I}_{d}$, $\hat{\boldsymbol{\theta}} = \mathbf{0}_{d}$, $\mathbf{f} = \mathbf{0}_{d}$;\\
	\For{t = 2, ..., $\mathcal{T}$}{
	\vspace{0.1cm}
	Sense interference vector $\mathbf{\hat{s}}_{t} = [s_{1},...s_{S}]$;\vspace{0.2cm}
	
	Create constrained action space $\mathcal{W'} = \{ w_{i} \in \mathcal{W}: \mathcal{D}(w_{i}|\mathcal{F}_{t-1}) < \hat{d} \}$; \vspace{0.2cm}
	
	Using $\mathbf{\hat{s}}_{t}$ and $\mathcal{F}_{t-1}$ assemble context vectors $\mathbf{x}_{w_{i},t} = [\xi_{1},...,\xi_{d}], \; \; \forall \; w_{i} \in \mathcal{W'}$;\vspace{0.2cm}
	
	Sample $\tilde{\boldsymbol{\theta}} \sim \mathcal{N}(\hat{\boldsymbol{\theta}}, \mathbf{B}^{-1})$;\vspace{0.2cm}
	
	Select LFM waveform $\mathbf{w}_{i}(t) = \argmin_{w_{i} \in \mathcal{W'}} \langle \mathbf{x}_{w_{i},t}, \tilde{\boldsymbol{\theta}} \rangle$;\vspace{0.2cm}
	
	Observe cost $\mathcal{C}(w_{t}, \mathbf{s}_{t})$; \vspace{0.2cm}
		
	Update distribution parameters $\mathbf{B} = \mathbf{B} + \mathbf{x}_{w_{i},t} \mathbf{x}_{w_{i},t}^{T}$, $\mathbf{f} = \mathbf{f} + \mathbf{x}_{w_{i},t} \mathcal{C}(w_{t},\mathbf{s}_{t})$, and $\boldsymbol{\hat{\theta}}_{t} =\mathbf{B}^{-1} \mathbf{f}$;
	}
\end{algorithm}	

\begin{algorithm}[t]
	\setlength{\textfloatsep}{0pt}
	\label{algo:exp3}
	\caption{Constrained Linear Contextual EXP3}
	\SetAlgoLined
	Initialize learning rate $\varepsilon \in (0,1)$, exploration distribution $\pi$, and exploration parameter $\gamma \in [0,1]$\\
	\For{t = 2,...,$\mathcal{T}$}{
		\vspace{0.1cm}
		Sense interference vector $\mathbf{\hat{s}}_{t} = [s_{1},...s_{S}]$;\vspace{0.2cm}		
		
		Create constrained action space $\mathcal{W'} = \{ w_{i}  \in \mathcal{W}: \mathcal{D}(w_{i}|\mathcal{F}_{t-1}) < \hat{d} \}$; \vspace{0.2cm}
		
		Using $\mathbf{\hat{s}}_{t}$ and $\mathcal{F}_{t-1}$ assemble context vectors $\mathbf{x}_{w_{i},t} = [\xi_{1},...,\xi_{d}], \; \; \forall \; w_{i} \in \mathcal{W'}$;\vspace{0.2cm}
		
		For each $w_{i} \in \mathcal{W'}$ \vspace{.1cm} set $P_{t}(w_{i}) \leftarrow \gamma \pi(w_{i})+(1-\gamma) \frac{\exp \left(-\varepsilon \sum_{j=1}^{t-1} \hat{\mathcal{C}}_{j}(w_{i}, \mathbf{s}_{j})\right)}{\sum_{w_{i}^{\prime} \in \mathcal{W'}} \exp \left(-\varepsilon \sum_{j=1}^{t-1} \hat{\mathcal{C}}_{j}\left(w_{i}^{\prime}, \mathbf{s}_{j} \right)\right)}$; \vspace{0.2cm}
		
		Sample $w_{t} \sim P_{t}$ and observe $\mathcal{C}(w_{t},\mathbf{s}_{t})$; \vspace{.2cm}
		
		Set $\boldsymbol{\hat{\theta}}_{t} \leftarrow \mathbf{Q}_{t}^{-1} \mathbf{x}_{w_{i},t} \mathcal{C}_{t}$ and $\hat{\mathcal{C}}_{t}(w_{i},\mathbf{s}_{t}) \leftarrow \langle \mathbf{x}_{w_{i},t}, \boldsymbol{\hat{\theta}}_{t} \rangle$;						
	}	
\end{algorithm}

TS simply involves selecting actions based on the posterior probability that they are optimal. The posterior distribution $\mathbb{P}(\boldsymbol{\theta}|\mathcal{F}_{t-1})$ is computed using Bayes' rule and a randomly initialized normal prior\footnote{The implementation in this paper considers Gaussian Thompson Sampling, which exploits the normal-normal conjugacy property to yield a normally distributed posterior from which samples can be efficiently generated.}. 

Since a variety of actions may be selected by the algorithm, we also wish to mitigate distortion effects by constraining the set of possible waveforms. Thus, at each step we solve the constrained optimization problem
\begin{equation}
\begin{aligned}
& \underset{w_{i} \in \mathcal{W}}{\text{minimize}}
& & \E[\mathcal{C}(w_{i},\mathbf{s}_{t})| \mathbf{\hat{s}}_{t}, \mathcal{F}_{t-1}] \\
& \text{subject to}
& & \mathcal{D}(w_{i}|\mathcal{F}_{t-1}) < \hat{d},
\end{aligned}
\label{eq:const}
\end{equation}
where $\E[C({w_{i}},\mathbf{s}_{t})|\mathbf{\hat{s}}_{t}, \mathcal{F}_{t-1})]$ is the posterior expected cost of transmitting $w_{i}$ based on the information observed until the previous PRI and $\hat{d}$ is a tolerable level of distortion.
\begin{figure*}[t!]
	\centering
	\begin{subfigure}[]
		\centering
		\includegraphics[scale=0.40]{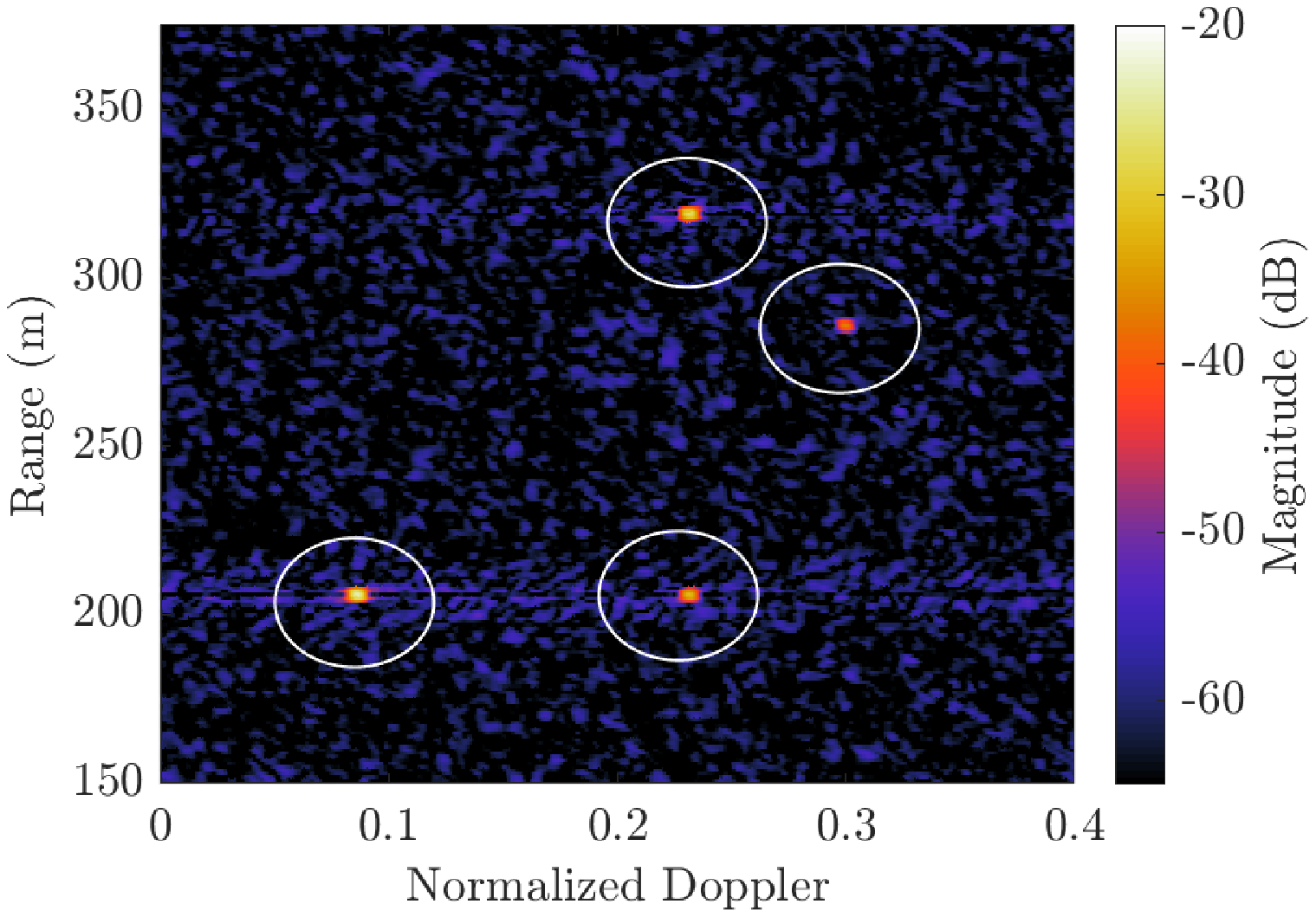}
	\end{subfigure}
	\begin{subfigure}[]
		\centering
		\includegraphics[scale=0.40]{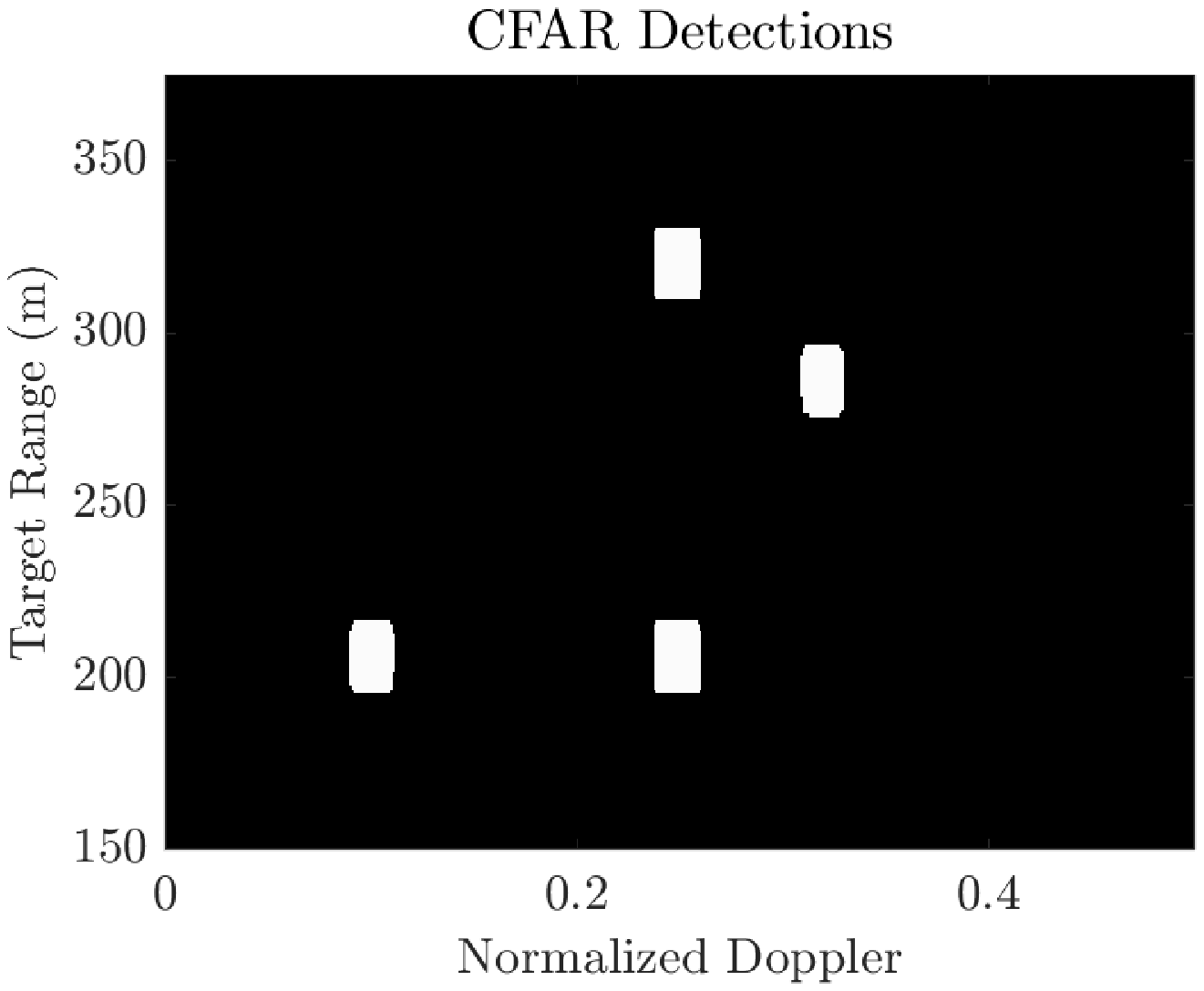}
	\end{subfigure}
	\begin{subfigure}[]
		\centering
		\includegraphics[scale=0.40]{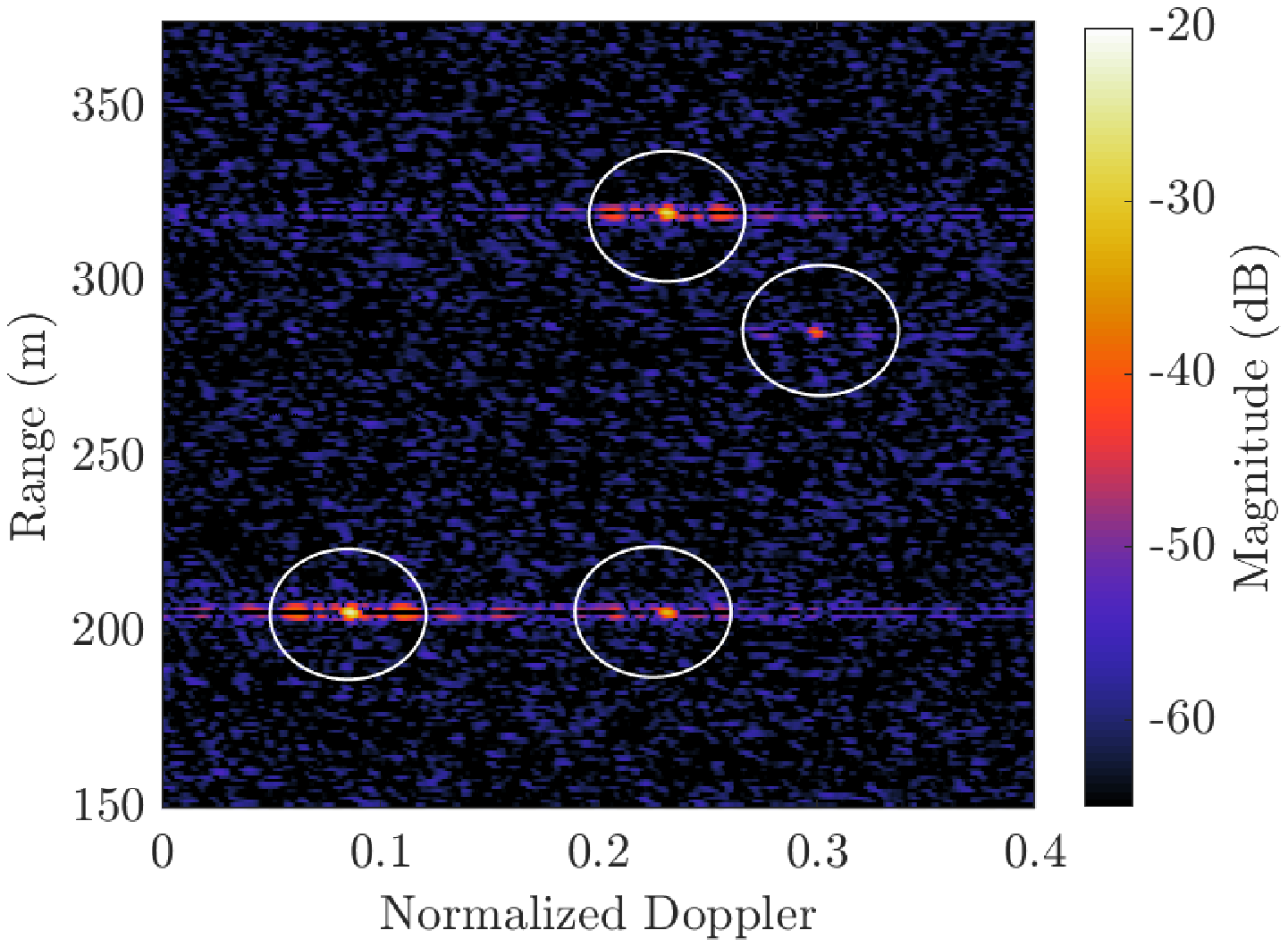}
	\end{subfigure}
	\begin{subfigure}[]
		\centering
		\includegraphics[scale=0.40]{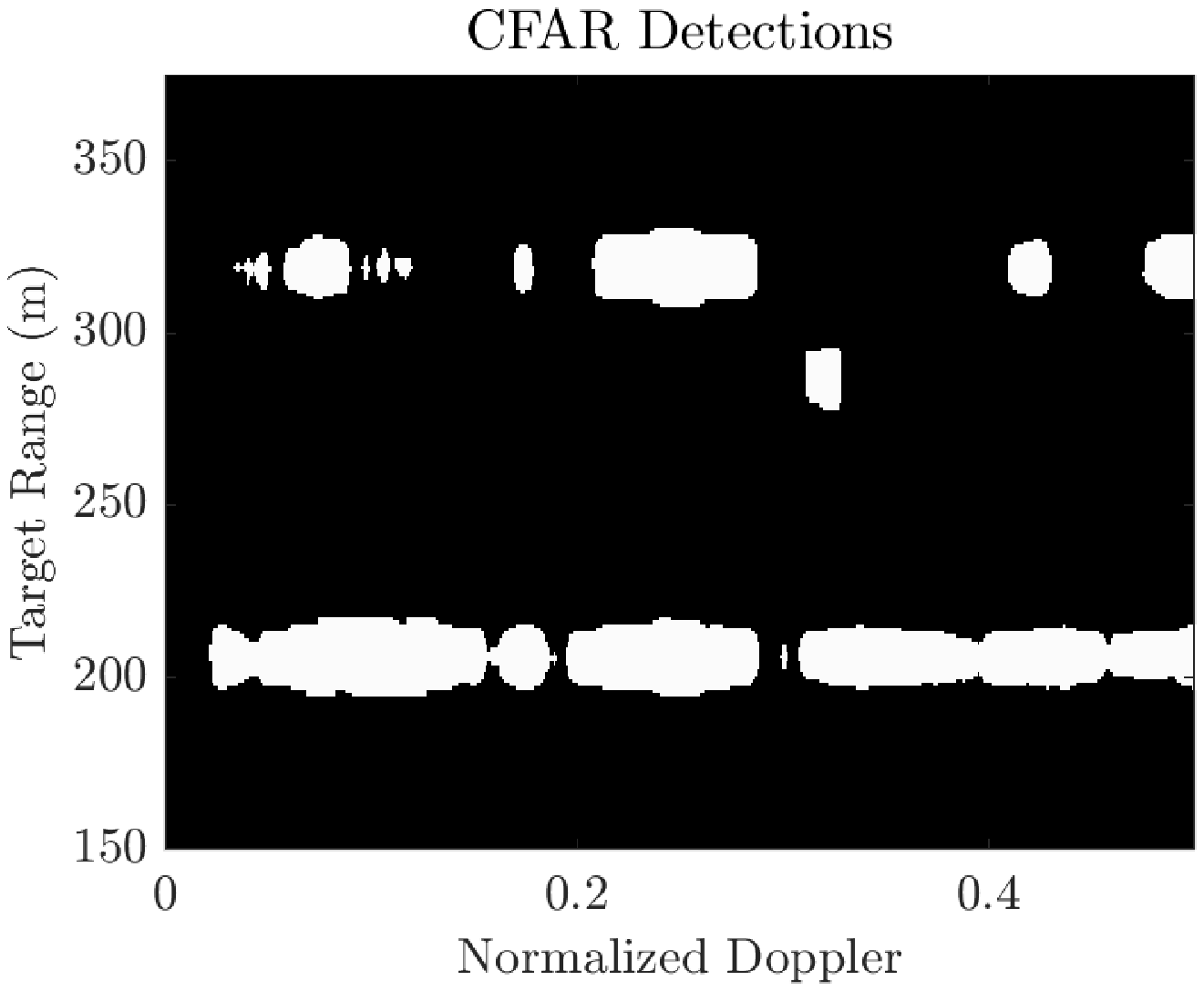}
	\end{subfigure}
	\caption{(a) Range-Doppler map of CR following constrained TS. (b) CFAR detections of CR following constrained TS for a desired $P_{\texttt{fa}} = 10^{-6}$. Note that the four true targets are detected with no false alarms. (c) range-Doppler map of CR following unconstrained TS. (d) CFAR detections of CR following unconstrained TS for a desired $P_{\texttt{fa}} = 10^{-6}$. Note that distortion effects cause many false alarms as energy spreads across the Doppler domain.}
	\label{fig:conUn}
	\vspace{-.2cm}
\end{figure*}

The full procedure of building the posterior distribution and obtaining samples can be seen in Algorithm \ref{algo:ts}. We note that the TS algorithm is computationally efficient as long as (\ref{eq:const}) is easily solvable. This is the case when the inner product $\langle \boldsymbol{\theta}, \mathbf{x}_{w_{i},t} \rangle$ can be computed and a sample can be efficiently generated from $\mathbb{P}(\boldsymbol{\theta}|\mathcal{F}_{t-1})$. Since the radar's waveform catalog is of finite cardinality $\lvert \mathcal{W} \rvert = W < \infty$ and the action set is further constrained such that $\lvert \mathcal{W'} \rvert \leq \lvert \mathcal{W} \rvert$ each PRI, the TS algorithm is very efficient in most practical cases.

\subsection{Adversarial Linear Contextual Bandits and the EXP3 Algorithm}
In addition to the stochastic linear contextual bandit model, we also consider the \textit{adversarial} linear contextual bandit framework, which makes the gentler assumption that costs are arbitrarily selected by an intelligent adversary \cite{lattimore}. This is a more general setting than the stochastic bandit which relaxes the assumption that costs are drawn from a fixed model $\langle \boldsymbol{\theta}, \mathbf{x}_{w_{i},t} \rangle + \eta_{t}$. The cost structure in the adversarial linear contextual bandit model is expressed by
\begin{equation}
\mathcal{C}(w_{i}, \mathbf{s}_{t}) = \langle \boldsymbol{\theta}_{t}, \mathbf{x}_{w_{i},t} \rangle,
\end{equation}
where the parameter vector $\boldsymbol{\theta}_{t} \in \mathbb{R}^{d}$ is now a time-varying quantity and the radar must learn a model which is nonstationary in general. The only source of randomness in the radar's regret is the distribution of the waveforms the radar transmits, which may reflect scenarios when the radar's waveform impacts the channel quality, such as in the presence of an intelligent jammer.

To balance exploration and exploitation in this setting, the radar uses a constrained variant of the EXP3 algorithm, first introduced by Auer in \cite{exp3} and widely used in adversarial bandit problems thereafter. EXP3 uses \textit{exponentially weighted estimation} to approximate the expected cost of each waveform before transmission. The exponentially weighted estimator $P_{t}: \mathcal{W'} \mapsto [0,1]$ is given by the probability mass function
\begin{equation}
\tilde{P}_{t}(w_{i}) \propto \exp \left(\varepsilon \sum_{j=1}^{t-1} \hat{\mathcal{C}}_{j}(w_{i},\mathbf{s}_{j})\right)
\label{eq:expweight}
\end{equation}
where $\varepsilon \in (0,1)$ is the learning rate and $\hat{\mathcal{C}}_{j}$ is an estimate of the cost at PRI $j$. To control the variance of the cost estimates, (\ref{eq:expweight}) is mixed with an arbitrary exploration distribution $\pi: \mathcal{W} \mapsto [0,1]$ where $\sum_{w_{i} \in \mathcal{W'}} \pi(w_{i}) = 1$. In the constrained implementation, $\pi$ is set to be uniform over $\mathcal{W'}$ each PRI. The mixture distribution from which waveforms are selected is then given by
\begin{equation}
P_{t}(w_{i})=(1-\gamma) \tilde{P}_{t}(w_{i})+\gamma \pi(w_{i}),
\end{equation}
where $\gamma \in [0,1]$ is a mixing factor. Each PRI, the waveform is then sampled $w_{t} \sim P_{t}$. However, calculation of $P_{t}$ involves an estimation of the cost for each $w_{i} \in \mathcal{W'}$ at each time step. This is performed via least-squares. An estimate of the model is obtained using 
\begin{equation}
\boldsymbol{\hat{\theta}}_{t} = \mathbf{Q}_{t}^{-1} \mathbf{x}_{w_{i},t} \mathcal{C}_{t}, 
\end{equation}
where $\mathbf{Q}_{t} \in \mathbb{R}^{d \times d} = \sum_{w_{i} \in \mathcal{W'}} P_{t}(w_{i}) \mathbf{x}_{w_{i},t} \mathbf{x}_{w_{i},t}^{T}$. The cost estimate is then easily accessed through the inner product $\langle \mathbf{x}_{w_{i},t}, \boldsymbol{\hat{\theta}}_{t} \rangle$. 

EXP3 is nearly optimal in terms of worst-case regret, but the distribution of costs has a high variance \cite{lattimore}. However, through the bias-variance trade-off, this general model allows a CR to maintain acceptable performance in a wide range of environments. A description of the constrained EXP3 approach used here can be seen in Algorithm \ref{algo:exp3}.

\section{Simulation Study}
In this section, the proposed constrained online learning framework is evaluated in a radar-communications coexistence setting as well as in the presence of an adaptive jammer. In the former setting, the CR is the secondary user of a shared spectrum channel. The CR wishes to maximize its own detection performance by selecting waveforms which mitigate both interference and distortion effects in the processed data, while causing minimal harmful interference to other systems. The CR shares the channel with $N_{\texttt{BS}}$ cellular base stations (BSs), which are spatially distant from the radar. In the latter setting, the CR also wishes to maximize detection performance while mitigating distortion effects, but a single frequency-agile jammer is capable of tracking the radar's transmitted waveforms. Each setting is further described below.
\subsection{Coexistence Environment}
Each PRI, the radar selects a waveform $w_{i} \in \mathcal{W'}$ and observes $\mathcal{C}(w_{i},\mathbf{s}_{t})$. Once $M_{\texttt{CPI}}$ pulses are received, matched filtering and a 2D FFT are performed to create a range-Doppler map. Detection analysis is performed using a threshold selected by the 2D cell-averaging Constant False-Alarm Rate (CFAR) algorithm to evaluate the target detection properties of each CR scheme and traditional fixed band radar operation. Since the BSs are located far from the radar, small scale fading effects are absent and the interference channel is dominated by correlated shadowing. The aggregate interference at the radar is given by
\begin{equation}
\mathcal{I}_{\texttt{agg}} =  \textstyle \sum_{j = 1}^{N_{\texttt{act}}} P_{j} \mathcal{G}_{r} \left\lVert \mathbf{d}_{j} \right\rVert^{- \alpha} \exp({X_{j}}),      
\end{equation}
where $N_{\texttt{act}}$ is the number of active BSs, $P_{j}$ is the transmission power of BS $j$, $\mathcal{G}_{r}$ is the radar recieve antenna gain, $\mathbf{d}_{j}$ is the distance from BS $j$ to the radar, $\alpha$ is the path loss exponent, and $X_{j} \sim N(\hat{\mu}_{j}, \sigma^{2}_{j})$ is the data transmitted by BS $j$. The cellular network bandwidth is $20 \texttt{MHz}$ and BSs transmit between $40$ and $46.5 \texttt{dBm}$. The BSs are randomly distributed between $5$ and $6 \texttt{Km}$ from the radar. The path loss term is $\alpha = 3.5$. Each PRI, actions with $\mathcal{D}(w_{t} | \mathcal{F}_{t-1}) > (\hat{d} = 0.2)$ are eliminated.
\subsection{Adaptive Jamming Environment}
In this environment, the radar must avoid interference from an adaptive jammer. If the radar selects the same waveform for two consecutive PRIs, $w_{t} = w_{t-1}$, then during the next PRI a high-powered interfering signal occupies the bandwidth utilized by $w_{t}$. If the radar adapts its waveform, $w_{t} \neq w_{t-1}$, the jammer occupies the bandwidth it utilized in the previous PRI. This is a more challenging scenario for the CR than the coexistence environment as the radar's actions influence the behavior of the interference and the jammer may adapt every PRI. In the following simulation, we consider a constant jammer-to-noise-ratio (JNR) of $20 \texttt{dB}$ at the radar when interference occurs. Once again, actions with $\mathcal{D}(w_{t} | \mathcal{F}_{t-1}) > (\hat{d} = 0.2)$ are eliminated.\\

\subsection{Simulation Results}
In these numerical examples, we consider $M_{\texttt{CPI}} = 400$ pulses and a PRI of $.1024 \texttt{ms}$. The waveform catalog $\mathcal{W}$ consists of 55 elements, which range in bandwidth from $10$ to $100 \texttt{MHz}$ and baseband center frequency from $-45$ to $45 \texttt{MHz}$. Detection performance is examined during the exploration phase, where pulse-agility is utilized.

Figure \ref{fig:conUn} shows the range-Doppler response and CFAR based threshold detections of the CR following the constrained TS approach of Algorithm \ref{algo:ts} and an unconstrained approach where $\mathcal{W'} = \mathcal{W}$ each PRI. The constrained approach results in lower sidelobe levels, and energy returns are concentrated around the true target locations. For a desired $P_{\texttt{fa}} = 10^{-6}$, we note that constrained TS yields no false alarms while unconstrained TS detects many false targets due to the spreading of energy across the Doppler domain.
\begin{figure}[t]
	\centering
	\includegraphics[scale=0.6]{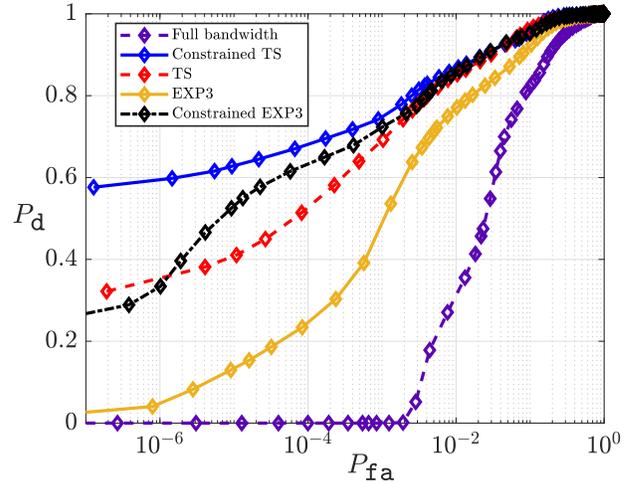}
	\caption{Empirical ROC using matched filter and CFAR detection analysis. Coexistence scenario with cellular interference $T_{c}$ = 7 pulses. Each data point corresponds to 30 algorithm runs, each consisting of 25 CPIs. Each CPI consists of 400 pulses.}
	\label{fig:tc7}
\end{figure}

Figure \ref{fig:tc7} shows Receiver Operating Characteristic (ROC) curves for each of the constrained and unconstrained online learning algorithms in the coexistence environment. Performance is compared to traditional radar operation utilizes the entire bandwidth of the shared channel. The cellular interference has a coherence time $T_{c}$ of 7 PRI's, meaning that the interference will remain constant for at least that amount of time. In this setting, Constrained TS provides the best overall performance, managing positive detection rates of above $0.5$ at false alarm rates less than $10^{-5}$. Unconstrained TS provides reasonable performance, but is significantly worse than its constrained counterpart due to higher sidelobe levels from waveform adaptations. The adversarial bandit algorithms converge slower than the stochastic bandits, as evidenced by worsened detection performance. This is because the i.i.d assumption associated with the stochastic model describes the coexistence scenario well. However, it should be noted that the constrained EXP3 approach provides a noticeable benefit over the unconstrained algorithm, especially at low false alarm rates.
\begin{figure}[t]
	\centering
	\includegraphics[scale=0.6]{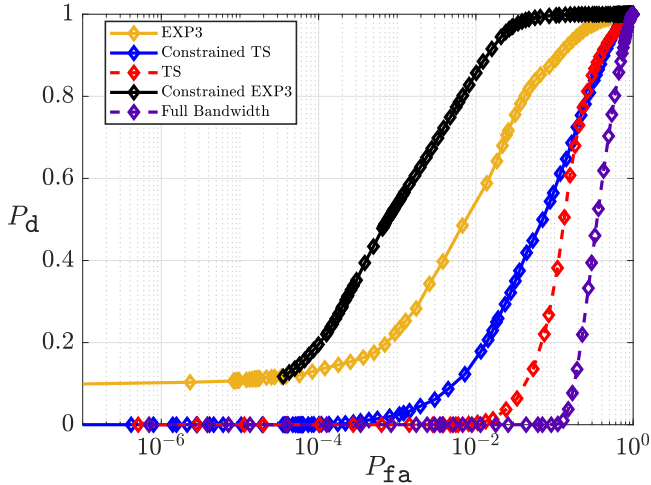}
	\caption{Empirical ROC using matched filter and CFAR detection analysis. Adaptive jammer scenario, with potential to adapt each PRI. Each data point corresponds to 30 algorithm runs, each consisting of 25 CPIs. Each CPI consists of 400 pulses.}
	\label{fig:tc3}
\end{figure}

In Figure \ref{fig:tc3}, detection performance in the adaptive jamming scenario is shown. In this case, the TS algorithms under the stochastic bandit model performs poorly compared to the EXP3 algorithms under the adversarial model, which is consistent with what can be expected given the assumptions of each model. In this scenario, the improvement from using the constrained waveform catalog is pronounced even at high values of $P_{\texttt{fa}}$. This is to be expected as the radar is varying its waveform more frequently than in the coexistence environment due to the adaptivity of the jammer. Thus, we note that the constrained learning approach may be particularly effective in scenarios for which the interference is changing rapidly and unpredictably.

\section{Conclusion}
A constrained online learning approach for pulse-agile cognitive radar was presented. This structure can be applied to a wide range of decision-making algorithms\footnote{While this study has focused on contextual bandit algorithms, this learning approach was also tested in the less general multi-armed bandit setting with similar results for Thompson Sampling and EXP3.}, demonstrated by the constrained linear TS and EXP3 algorithms presented in Algorithms \ref{algo:ts} and \ref{algo:exp3}. Through simulations in dynamic radar-communication coexistence settings, the proposed scheme was demonstrated to reduce distortion effects for favorable detection performance when a cost function based on interference avoidance is used. This computationally feasible scheme also has the potential to be used in tandem with previously proposed adaptive processing techniques such as those described in \cite{richardson, nimpc, ravenscroft, scholnik11}. Open problems include analytically characterizing the performance of the algorithms with respect to the behavior of the environment and a detailed study of the trade-off between interference mitigation and distortion effects when utilizing pulse-agility. Future work could also focus on characterizing the effect on tracking performance or extending this learning approach to distributed sensing applications involving multiple pulse-agile radars.

\section{Acknowledgment}
The authors would like to thank Benjamin Kirk for helpful discussion and aid in formulating the simulations.

\bibliographystyle{IEEEtran}
\bibliography{bibli}

\appendix
\subsection{Proof of Remark 1}
Since $\mathcal{C}(w_{i})$ is composed of three terms, $\beta_{1} \texttt{BW}_{c}(w_{i},\mathbf{s}_{t})$, $\beta_{2} \texttt{BW}_{miss}(w_{i},\mathbf{s}_{t})$, and $\beta_{3}\mathcal{D}(w_{i}|\mathcal{F}_{t-1})$, it is sufficient to show each term is Lipschitz continuous for any pair of waveforms $(w_{i}, w_{j}) \in \mathcal{W} \times \mathcal{W}$. The first term considers the bandwidth shared with a fixed interference vector $\mathbf{s}_{t}$. Let the Lipschitz metric be $\mathcal{L}(w_{i},w_{j}) \triangleq L_{1} \lvert \Delta f_{c} \rvert + L_{2} \lvert \Delta \texttt{BW} \rvert$ where $\Delta f_{c}$ is the difference in center frequency between $w_{i}$ and $w_{j}$, $\Delta \texttt{BW}$ is the difference in bandwidth between $w_{i}$ and $w_{j}$, and $L_{1}, L_{2} > 0$ are fixed constants. Thus, waveforms with $\mathcal{L} \rightarrow 0$ will yield similar values of $\texttt{BW}_{c}$ due to spectral overlap with some $\mathbf{s}_{t}$. The second term is also dependent on the frequency content of each waveform, so a similar argument holds for $\texttt{BW}_{miss}$. Finally, the third term is dependent on the distance in $\texttt{BW}$ and $f_{c}$ from a fixed waveform $w_{t-1}$. Due to the similar structure of $\mathcal{D}$ and $\mathcal{L}$, cases where $\mathcal{L}(w_{i},w_{j}) \rightarrow 0$ imply $\mathcal{D}(w_{i}|\mathcal{F}_{t-1}) \rightarrow \mathcal{D}(w_{j}|\mathcal{F}_{t-1})$ \qed

\end{document}